\documentclass[10pt, conference, compsocconf]{IEEEtran}
\usepackage[utf8]{inputenc}
\usepackage{latexsym}
\usepackage{ngerman}
\renewcommand{\figurename}{Figure}
\renewcommand{\tablename}{Table}
\renewcommand{\abstractname}{Abstract}
\usepackage{graphicx}
\usepackage{array}
\usepackage{multirow}
\usepackage{hhline}		
\usepackage{subfigure}
\usepackage{mathptmx}
\usepackage{mathtools}
\usepackage{amsmath}
\usepackage{amsfonts}
\usepackage{amssymb}
\usepackage{amsbsy}
\usepackage{amsthm}
\let\labelindent\relax
\usepackage{enumitem} 
\usepackage{here} 
\usepackage[square,sort&compress,comma,numbers]{natbib}

\begin{document}
\title{\vspace{-10mm}Modeling the Location Selection of\\ Mirror Servers in Content Delivery Networks\vspace{-3mm}}
\author{
%\IEEEauthorblockN{Peter Hillmann}
\IEEEauthorblockN{Peter Hillmann, Tobias Uhlig, Gabi Dreo Rodosek, and Oliver Rose}
\IEEEauthorblockA{Universit\"at der Bundeswehr M\"unchen\\
Neubiberg, 85577, Germany\\
%peter.hillmann@unibw.de}
Email: \{peter.hillmann, tobias.uhlig, gabi.dreo, oliver.rose\}@unibw.de}
}
\maketitle

%3. Vazirani: Approximation Algorithms

\begin{abstract}
For a provider of a Content Delivery Network (CDN), the location selection of mirror servers is a complex optimization problem.
%
%Location selection of mirror servers is a complex optimization problem for a Content Delivery Network (CDN) provider.
%Generally, the objective is to place the nodes centralized such that all customers get adequate service for their requests.
Generally, the objective is to place the nodes centralized such that all customers have convenient access to the service according to their demands. % Therefore, the location selection requires automated solution methods.
It is an instance of the \mbox{k-center} problem, which is proven to be NP-hard. Determining reasonable server locations directly influences run time effects and future service costs. We model, simulate, and optimize the properties of a content delivery network. Specifically, considering the server locations in a network infrastructure with prioritized customers and weighted connections.
A simulation model for the servers is necessary to analyze the caching behavior in accordance to the targeted customer requests. We analyze the problem and compare different optimization strategies. For our simulation, we employ various realistic scenarios and evaluate several performance indicators. Our new optimization approach shows a significant improvement. The presented results are generally applicable to other domains with k-center problems, e.g., the placement of military bases, the planning and placement of facility locations, or data mining.
\end{abstract}


\IEEEpeerreviewmaketitle

%\noindent\begin{keywords}
%Content Delivery Networks, k-center problem, facility location problem, weighted network infrastructure, mirror server placement, data mining, clustering
%\end{keywords}
\vspace{-2mm}
\section{INTRODUCTION}\label{introduction}
Content Delivery Networks (CDN) are large distributed systems of servers, which provide a caching infrastructure. To face the rapidly growing demand of content, CDN providers offer large data capacity within distributed storage infrastructures across the Internet. It is an efficient and common method to provide content to end-users providing high availability and high performance. The mirror servers copy and cache the content of an original source. Locations of mirror servers have to be determined with a short distance to customers with a view to provide adequate service and convenient access. Customers request content from a website and the according answer should be returned by the server within the shortest distance to speed up the responsiveness. Usually, the implementation is transparent to the user. This improves the Quality-of-Experience (QoE) for a customer as well as reducing transfer time to requests and load times for content. To ensure this, the corresponding content has to be available on the caching server. In most cases, the stored content is from contracting companies and offloaded content of Internet service providers (ISP). Nevertheless, a single server can not provide all the requested content immediately. If content is not available, then it is fetched from another mirror server or the original source. Furthermore, a cache replacement strategy has to decide which content is stored on the server. Large CDN providers like \mbox{Akamai%\cite{Akamai}
}, \mbox{Limelight%\cite{Limelight}
}, and Level 3 \mbox{Communications%\cite{level3}
} can handle enormous amounts of requests. Beside an improved service quality for the customer, advantages are decreased network load and thereby reduced transmission costs for an ISP. 
%\cite{Jiang09}

%improve the Quality-of-Experience (QoE) of the user.
%transparent

\begin{figure}[htb]
{
%\hspace*{-0.1cm}
\setlength{\abovecaptionskip}{0ex}%
\setlength{\belowcaptionskip}{0ex}%
\centering
\includegraphics[width=0.46\textwidth]{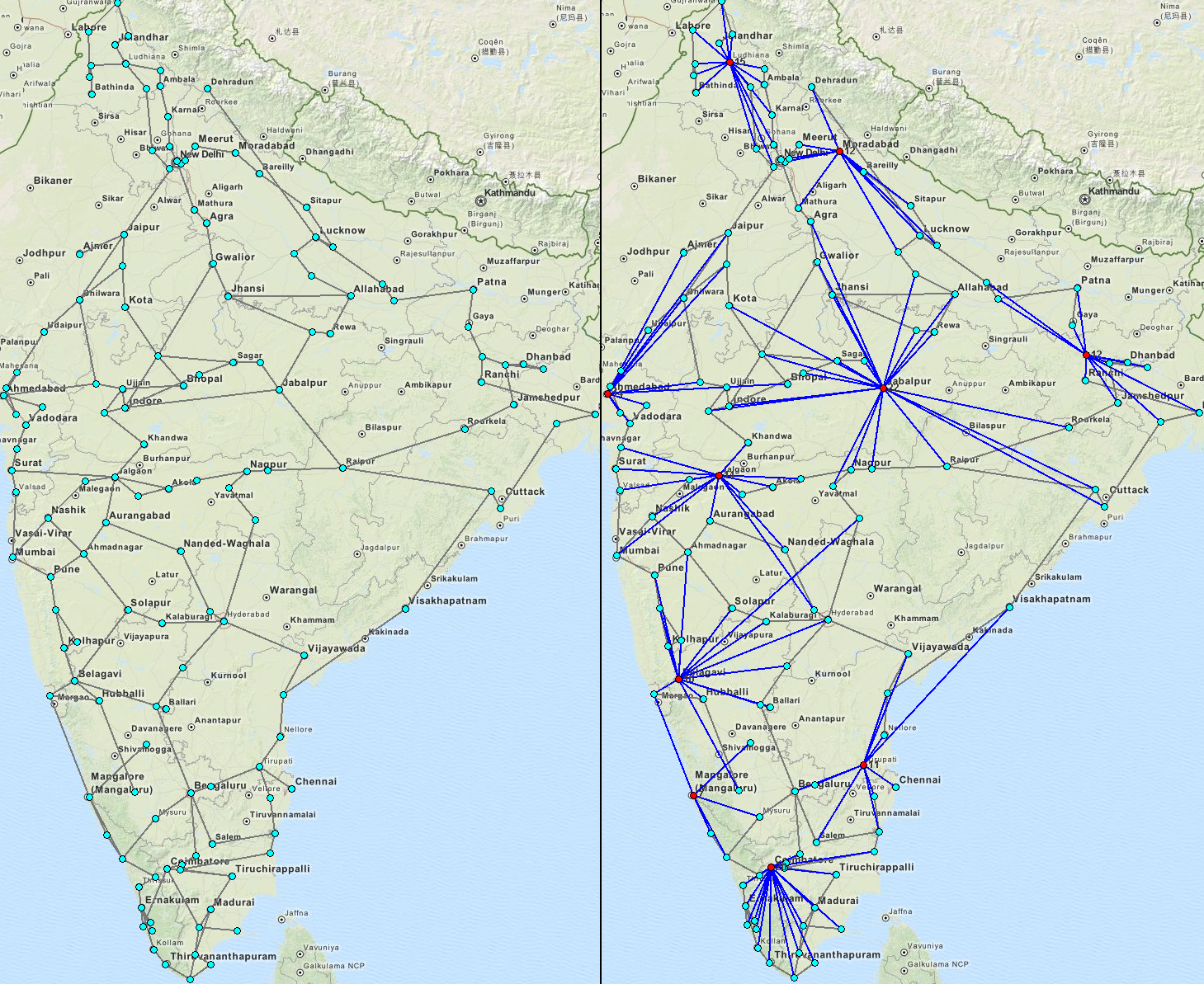}
%\vspace{-1mm}
\caption{Example of a scenario for the placement of mirror servers in India.}
\label{geographical}
\vspace{-3mm}
}
\end{figure}

%Generally, content should be stored close to the consumer to reduce access latency and foster continuous data transmission.
To that end, a predefined amount of mirror servers has to be placed and connected strategically to the ISP network infrastructure. 
An example of the problem is shown in \mbox{Figure \ref{geographical}.} The left side presents the geographical location of the network nodes and the edges of an infrastructure \cite{Knight2011}. % as well as their priorities and weights.
The right side includes the additionally registered mirror servers. The blue lines represent the assignments of the network nodes to their closest server and thereby the intended customer allocation.
\vspace{-2mm}

%The CDN provider controls the geographical and infrastructural placement of their servers, the placement of the content and the cache replacement strategy, as well as the decision about which client requests are served from which server. 
In this work, we focus on the geographical and infrastructural placement of the servers enforcing the placement close to an already existing network node. We present a model of the problem and analyze the influence of different placement decisions.
We simulate and automatically optimize the placement of the mirror servers and the assignment of customers to their most appropriate server with fitting content. %Thereby, we avoid requests from a customer to a mirror server which cannot provide the content immediately.
The assignment has a strong influence on the Cache-Miss-Ratio as well as access latency and network load.
To this end, we use simulation based optimization to analyze the necessary amount of mirror servers for an effective supply with respect to customer demands. 
Preliminary results on the caching behavior are discussed.
%cache replacement strategies with respect to customer demands and the network load.
%The location selection is a difficult management process during a planning phase. 
Our results support the management in its difficult decision process during a planning phase.
The underlying problem is of importance in other application areas. For example in logistics, the placement of warehouses, fire stations and hospitals, as well as other facilities are based on the similar k-center problem.

Our paper has the following structure. After the introduction in Section I, we describe a typical scenario of our application area in Section II. We briefly discuss the complexity of the problem and introduce related optimization techniques in the Sections III and IV. The description of our model and the novel optimization algorithm are explained in Section V. The evaluation and assessment are presented in Section VI. The reference algorithms are outlined in Section VII. We summarize the paper and discuss future work in Section VIII.\\

%\vspace{1mm}
\vspace*{-3mm}
\section{SCENARIO}\label{scenario}
\begin{figure}[hbtp]
{
\setlength{\abovecaptionskip}{0ex}%
\setlength{\belowcaptionskip}{0ex}%
\centering
\includegraphics[width=0.37 \textwidth]{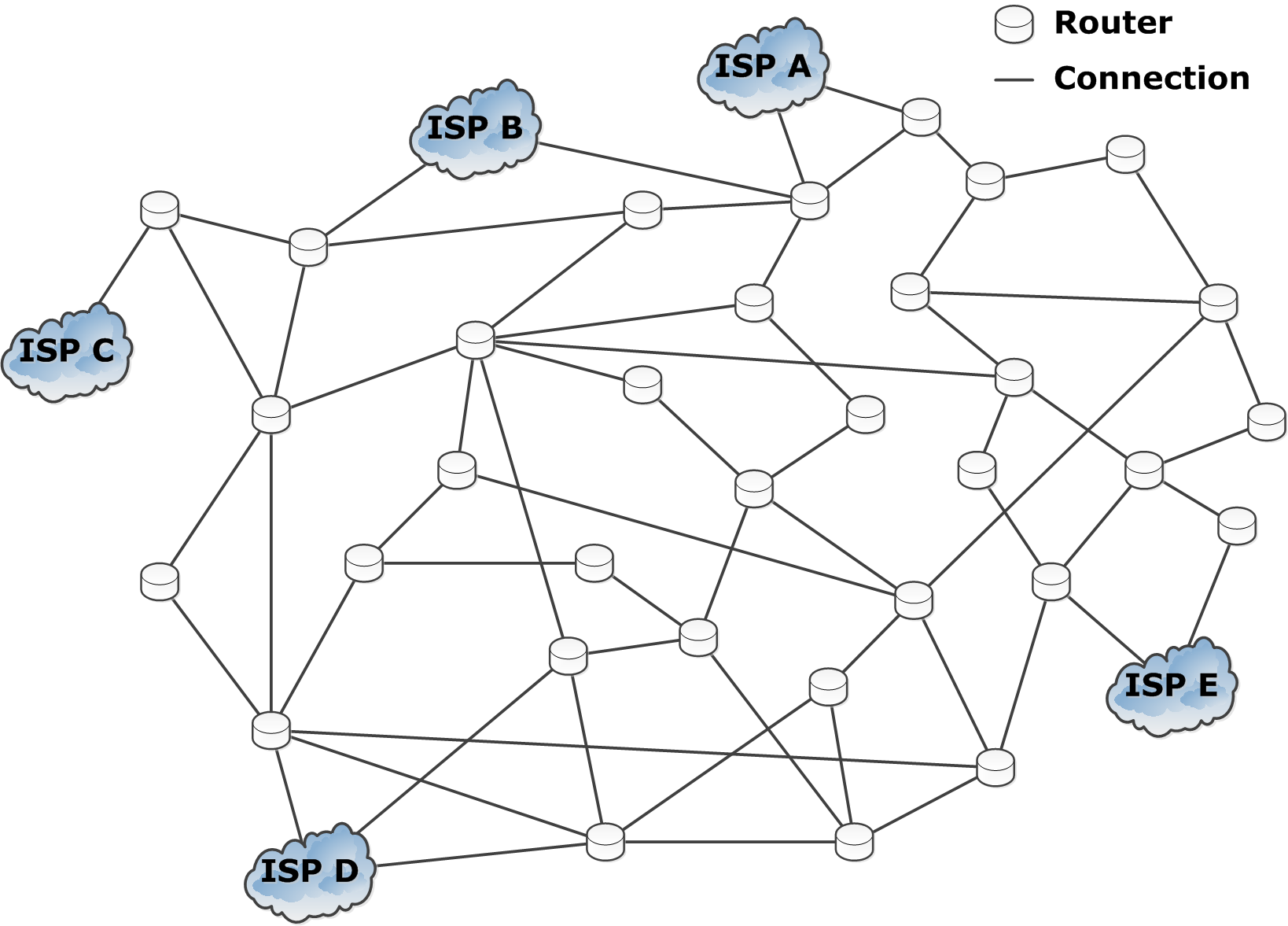}
\caption{Abstract view of a network infrastructure from an ISP.} %Example Internet backbone network topology in Europe \cite{Knight2011}, non-hierarchical infrastructure.}
\label{fig:backboneTopology}
\vspace*{-0.2cm}
}
\end{figure}

Consider the following scenario: a CDN provider intends to expand its business and wants to set up a completely new storage infrastructure in a country. The first step of the planning process is to analyze the network infrastructure and the connected users. Figure \ref{fig:backboneTopology} shows an abstract example of a typical infrastructure which is non-hierarchical. %Because of the high-performance connection, we focus on mirror servers placed within the backbone network topology.
To reduce complexity, multiple customers connected to the same gateway or network node are merged to a single node. Different approaches to generate the map are described in Section \ref{relatedwork}. 

\vspace*{-0.3cm}

The CDN provider tries to place the mirror servers close to the customers. %, to reduce access latency and network load.
In this case, \glqq close\grqq{} refers to the network infrastructure and not to the geographical location. The management has to decide where to place the mirror servers geographically as well as structurally and organize the assignment of the customers to a server. Furthermore, the distances have to be balanced against the correlated customer demands. 
The various customer demands are summarized in a profile. 
Customers send requests for specific content according to their profile to the server. This influences the dynamic caching behavior and the currently available content on the server. The requested content on a server should be similar to increase the cache hit ratio. This lowers the amount of requests from a mirror server to the original source, lowers the network utilization and reduces access latency.
For the management of a CDN provider, it is important to know the necessary quantity of mirror servers for a given scenario to achieve a predefined objective. An example of such an optimization problem is shown in \mbox{Figure \ref{geographical}}.

%The placement of the mirror servers and the assignment of customers to their closest server have to be simulated and optimized automatically. An example is shown in Figure \ref{geographical}. The left side presents the geographical location of the network nodes, the edges between as well as their priorities and weights. The right side includes the additionally registered mirror servers. The lines represent the assignments of the nodes to their closest server and thereby the customer allocation.
%\vspace{1mm}
%\begin{figure}[htb]
%{
%\centering
%\includegraphics[width=0.45\textwidth]{fig/scenarioBoth.png}
%\caption{Scenario for placement of Distribution Center.}
%\label{geographical}
%}
%\end{figure}

Due to the complexity of the challenge, the placement of mirror servers and the assignment of customers to a server have to be technically supported with simulation and optimization. According to the scenario and various application areas, we need to answer the following important questions:

\begin{itemize} %[itemsep=2pt] 
\item How to model the system capturing its distinct properties, e.g. customer priorities and connection bandwidth?
\item Where are mirror servers to be placed in the infrastructure in order to obtain short and efficient transmission paths?
\item Which customer is assigned to which mirror server?
\item How large is the influence of the positively correlated customer demands for a dedicated mirror server on the caching behavior?\\ % (modeling of customer profiles)
\end{itemize}

\vspace*{-3mm}
\section{PROBLEM DESCRIPTION}
The scenario described in Section \ref{scenario} is based on the k-center problem. For a given set of locations $V$, a predefined number of central locations $K$ have to be determined. The k-center problem considers the minimization of the maximum distance between a location and its nearest center. % \cite{Chaudhuri1998}.
In our case, it is the maximum distance of a customer to its assigned mirror server. The problem is proven to be NP-hard \cite{Potikas09}. The problem can be specified using a strongly connected graph topology \mbox{$G(V,E)$} with vertices ($v_{ i } \in V{}\ with{}\ i :=\{1,..,n\}$) and edges ($e_{ p } \in E{}\ with{}\ p :=\{1,..,m\}$). The vertices have different priorities and the edges have different weight depending on their characteristics. The k-center problem is defined on a complete, undirected graph. The objective function $d$ defines the fitness value d($v_i$,$v_j$) for an edge e($v_{i}$,$v_{j}$) between two vertices ($v_{i}$,$v_{j}$), satisfying the triangle inequality. It selects the best edge from a vertex $v_{i}$ to one of the calculated locations of a center node ($k_{u} \in K{}\ with{}\ u :=\{1,..,l\}\ and{}\ l \leq n $). A center node $k_{u}$ can only be placed on a location of a vertex. Considering our domain, a mirror server has to be connected to an existing network node. The set of all fitness values $d$ from every vertex $v_{i}$ is defined as $D$. We intend to place the number of center nodes $K$ to minimize the maximum $d$($v_{i}$, $k_{u}$) from a vertex $v_{i}$ to its \mbox{best $k_{u}$.} The objective criterion is calculated using \mbox{Equation \ref{form:max}:}
%\vspace*{-0.2cm}
\begin{equation}
D_{center}(K) = \underset{ } { min }\text{ } \underset{v_i = 1, ..., n }{ max } \text{ } \underset{ k_u \in K } {min}\text{ } d(v_i, k_u)
\label{form:max}
\end{equation}
%\vspace*{-0.3cm}
%We intend to place k centers, with k being an integer $0 < k \leq \left\lvert V \right\rvert$, to minimize the maximum d(v, k) from a vertex to its \mbox{closest k:}

%In our case, we want to set up k mirror servers and minimize the maximum distance from a customer to its nearest mirror server.

%\input{content/requirements}
\vspace{1mm}
\section{RELATED WORK}\label{relatedwork}
There is a lot of existing work on geographical mirror server placement. Most research focuses on special hierarchical topologies. A line structure and a tree structure are analyzed in \cite{Carofiglio11a} and \cite{Krishnan00}. % Both approaches where adapted for this paper as benchmark. 
The work of \cite{Andreica09} and %, \cite{Khuller00},
\cite{Carofiglio11b} %and %\cite{Chankhunthod95}
propose solutions for the placement of storage servers in CDN with tree-like topologies. An approach for a ring-based architecture is described in \cite{Wang11}. However, these topologies do not reflect realistic infrastructures and the results of the specific cases cannot be adapted to the more general case.
There is less information about the placement of servers within an non-hierarchical or non-specialized infrastructure. \citet{Jamin00} analyzed different graph theoretic algorithms to determine the number and the placement of boxes for the purpose of network measurement with min k-center \cite{Vazirani01} and k-HST \cite{Bartal96}. But these works have little relation to the area of CDN. \citet{Rana09} proposes multiple heuristic approaches. An important and more general work is from \citet{Jamin01}. We use their proposed algorithms as benchmarks. However, either the aforementioned work covers our requirements only partially or the proposed approaches show only a modest performance.
Furthermore, these publications do not provide an adequate modeling of the problem. They use rough models without priorities and weighting and pay no attention to customer demands. Furthermore, these publications do not answer all the questions posed in Section \ref{scenario}.

Before the servers can be placed, the network infrastructure and the location of the customers need to be determined. %\citet{Tuncer13} highlight the importance of precise modeling of the network topology, but they do not provide a model.
% Several technologies exists To determine the network infrastructure and the location of the customers,.
Generally, the ISPs know the location of their customers and network nodes. If this information is not available one can obtain the information in an automated fashion. In \cite{Jafari13} an IP geolocation service is used to gather user locations in terms of latitude and longitude. Users are clustered based on their coordinates to build the network nodes. %Inet \cite{Winick02} is an autonomous system to generate Internet topologies. %Further possible approaches are traceroute \cite{Pignataro12} and traceback (\cite{Song01}, \cite{Savage00}, \cite{VINCENT10}) to generate the network infrastructure.
We assume that a corresponding scenario is given and we use the official data of \textit{the Internet Topology Zoo} \cite{Knight2011}.

%\vspace{-1mm}
\section{Modeling}
For the creation of an adapted model, the field of application is analyzed in more detail with specific characteristics. The entire system is strongly dependent on the underlying infrastructure. For an effective supply with the provided service, mirror servers should be connected strategically to the network infrastructure. Due to the typical expected high amount of requests from widely distributed customers, the servers need a high-performance connection with transit possibility.
%Because of the high-performance connection, we focus on mirror servers placed within the backbone network topology.
This is reached through a direct connection of the planned server to the backbone network % \cite{Wang2011}
 in the TIER 1 or 2 topology. These top-level infrastructures provide a high-performance accessibility, which makes them an ideal choice. Therefore, we focus on mirror servers placed within the backbone network topology. For our model, a server can only be connected to an existing network node. This corresponds to the placement of the server at this node. A placement of a server on a data link or in an area without direct connection option is seen as impractical. %So we do not create this possibility in our model.

With regard to customer modelling, our approach uses the input information from the underlying scenario. It includes the geographical location of the customers and their connection to the Internet gateway as well as the infrastructure.
% These many customers in an focused scenario are connected through the infrastructure.
To reduce the high amount of data for the modeling, we group multiple customers to a single network node. A group of customers connected to the same access node is aggregated with combined properties. This abstraction level is precise enough, because every Internet request from a customer has to pass this network node. Furthermore, the demands of the group of customers are aggregated to a group profile. To evaluate different groups of customers, the network node is assigned a priority. It is rated dependent on the represented amount of customers, their importance for the CDN provider, their payment, and frequency of requests.

Since we do not have real data of customer profiles from ISPs or CDN providers, pseudo realistic data is generated using the established model of Zipf-distribution \cite{Breslau1999}. To simulate the caching behavior each mirror server has been provided with a predefined amount of cache. The employed cache replacement strategy is  \textit{Least Frequently Used (LFU)}. %During simulation, every network node is repeatedly questioned whether a request is made to a server. Three different behaviors can occur: no request, a request matching the profile, and a new request not matching the profile.
During simulation, every group of customers sends repeatedly requests to their assigned server. Three different possibilities can occur during a repetition: no request, a request following the profile, or a novel request not according to the profile is send.

\begin{figure}[hbtp]
{
\setlength{\abovecaptionskip}{0ex}%
\setlength{\belowcaptionskip}{0ex}%
\centering
\includegraphics[width=0.46 \textwidth]{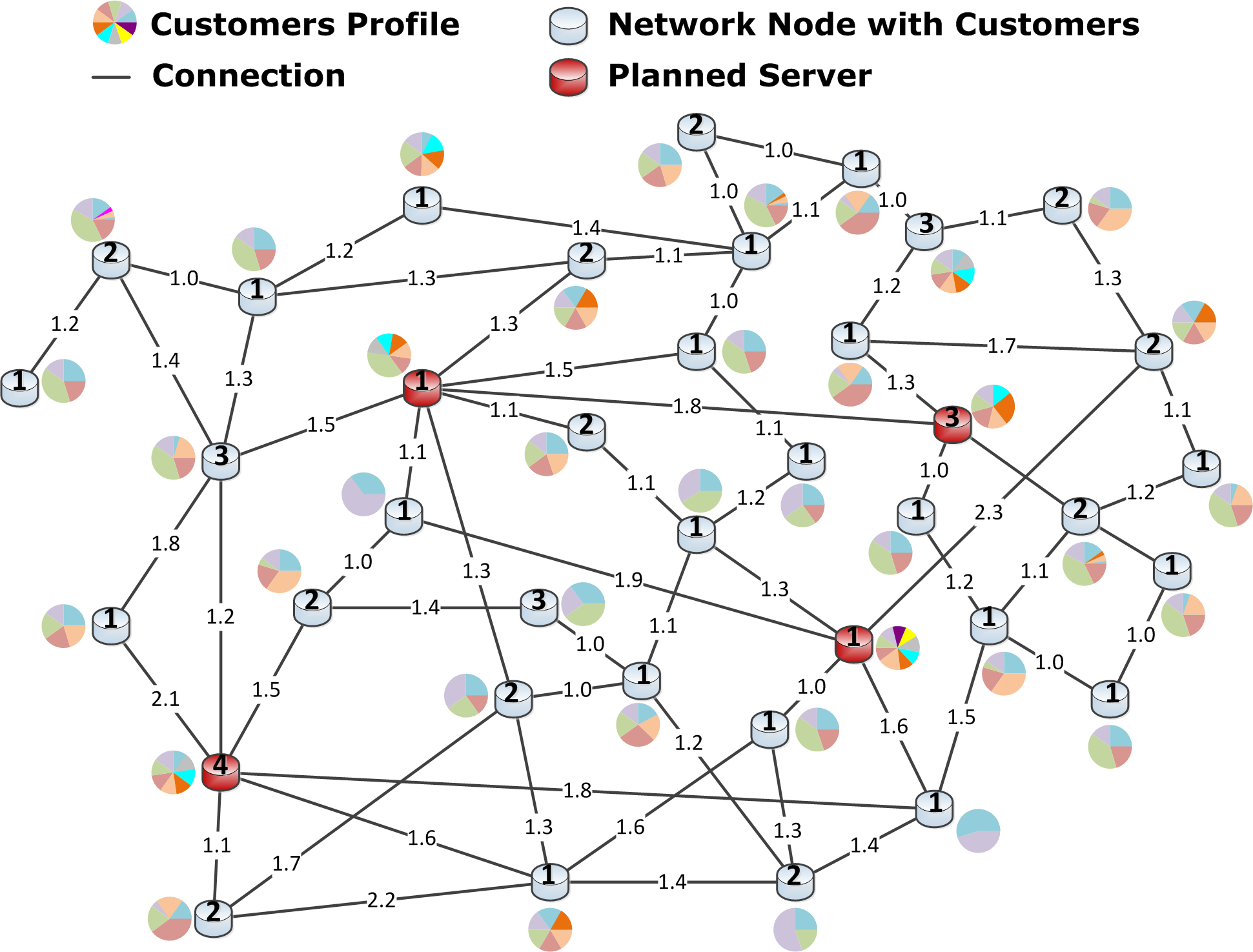}
\caption{Example of the abstract model with a non-hierarchical network infrastructure including weighted edges and grouped customers to a prioritized network node.}
\label{fig:AbstractScenario}
\vspace*{-0.2cm}
}
\end{figure}

These aspects result in an abstract model of our scenario, see Figure \ref{fig:AbstractScenario}. This includes prioritized network nodes for groups of customers and for placement possibilities of mirror servers. These network nodes are connected with each other via several data links using the existing infrastructure. The requested content is described by the profile of the network node, which is symbolized with a pie chart. %An example is presented in \mbox{Figure \ref{fig:AbstractScenario}.} 
\vspace*{-0.1cm}

For the optimization and the calculation of favorable server locations, it is mandatory to evaluate multiple locations with regard to the connection possibilities of assigned customers. 
% among the connections of a group of customers to the multiple planned locations of a server.
To calculate the distances between the network nodes, we use the metric of hops combined with the connection properties. A hop is a routing network component which is passed by a data packet during the communication. The delay of a transmission is mainly influenced by the processing of the network nodes themselves rather than by the physical length of a connection and the specific transmission medium. Therefor, all connections between the network nodes have a uniform virtual length of 1. The calculated distance between two nodes in the infrastructure results directly from the hop count. For the distance function and path finding method, we use Dijkstra's algorithm. % \cite{Dijkstr1959}.

\begin{figure}[hbtp]
{
\setlength{\abovecaptionskip}{0ex}%
\setlength{\belowcaptionskip}{0ex}%
\centering
\includegraphics[width=0.46 \textwidth]{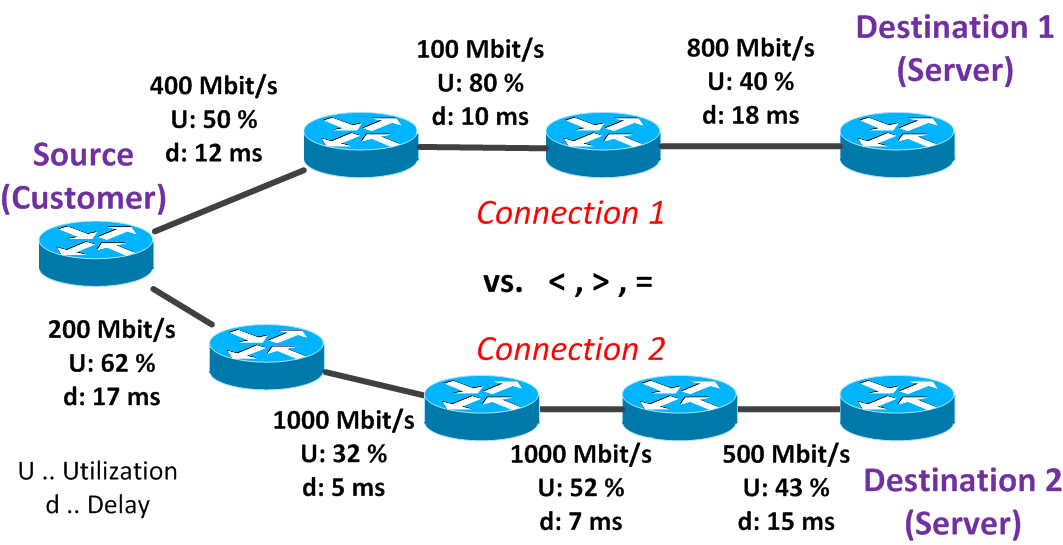}
\caption{Evaluation problem of different connections from a customer to several servers.}
\label{fig:PathComparission}
\vspace*{-0.3cm}
}
\end{figure}

Nevertheless, we have to pay attention to the specific properties of a data link to model the infrastructure more precisely. To compare different, shortest paths from a customer to several servers, a fitness value is necessary. Figure \ref{fig:PathComparission} illustrates the problem. We assign to every single edge in our model a specific weight, dependent on maximum bandwidth, average utilization, and mean delay. The necessary information is provided by the ISP. We map and combine this weight of an edge with the hop to a virtual length larger than 1. A worse link quality leads to a longer virtual length. A high-performance connection has a virtual length close to 1, which represents the best value in our model. The edge weight transformation to a virtual length still enables us to use Dijkstra's algorithm, since it calculates the shortest path based on distances. %Up to now, there exists no path finding method which deals directly with weighted edges and lengths.
%
%Nodes can have priorities to represent the relative importance of the connected users or other networks. Edges have different performance properties and get assigned a weight. Nevertheless, the connection between these servers is different.
\vspace*{-0.4cm}

The accumulated virtual path length from a customer to a server allows the comparison between different paths and enables efficient server assignments. To take the priority of a customer into account for the evaluation, the value of the path length is multiplied with the priority of the selected customer. %This represents the fitness value.
Dependent on the selected optimization criterion, we minimize the maximum distance value of a customer, the average distance value of all customers or other objectives.
%, which we optimize for the server placement

\begin{figure}[hbtp]
{
\setlength{\abovecaptionskip}{0ex}%
\setlength{\belowcaptionskip}{0ex}%
\centering
\includegraphics[width=0.46 \textwidth]{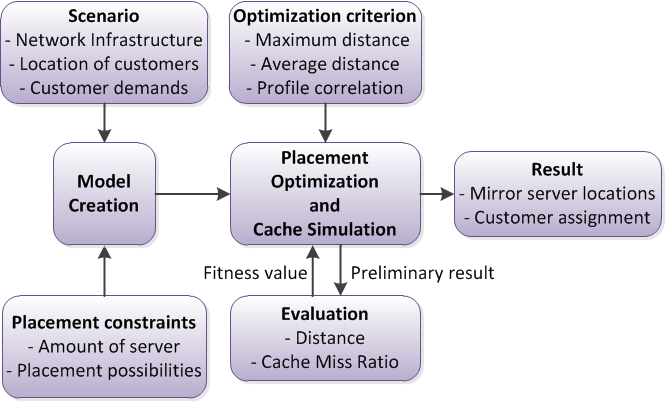}
\caption{Modeling approach to optimize the placement of mirror server.}
\label{fig:Model}
\vspace*{-0.2cm}
}
\end{figure}

Figure \ref{fig:Model} presents the model and the process of optimization. The data of the scenario is combined with the placement conditions to create an abstract model. The scenario contains the network infrastructure, locations of customers and their demands. The placement constraints include information on the specified amount of servers and the possible locations. Afterwards, the abstract model is simulated and the server locations are optimized. This is done with respect to the optimization criteria. It can be flexibly chosen to reflect the requirements of the management. In our case, we focus on a reduced network load with short transmission paths and high cache hit ratios. The process of optimization and simulation is iteratively repeated according to the used optimization algorithm. The result includes amongst other things the optimized server locations and the assignment of customers to their preferable server.
\vspace*{-0.2cm}

\begin{figure}[htb]
{
%\hspace*{-0.1cm}
\setlength{\abovecaptionskip}{0ex}%
\setlength{\belowcaptionskip}{0ex}%
\centering
\includegraphics[width=0.38\textwidth]{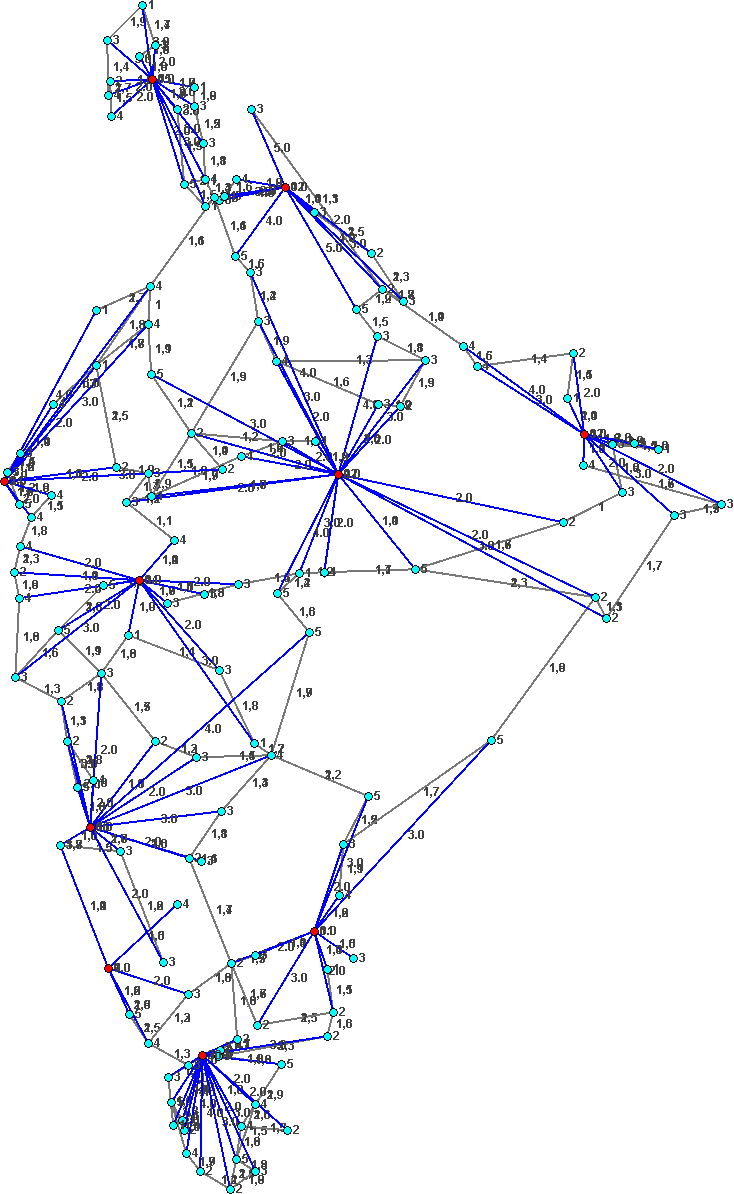}
\caption{Example model of the scenario in India and the planned mirror server including the assignment of requests.}
\label{fig:ExampleModel}
\vspace{-4mm}
}
\end{figure}

Figure \ref{fig:ExampleModel} shows the generated model and the preferred mirror server locations. It is optimized with our developed algorithm \textit{Dragoon} for shortest distance. The correlation of customer profiles is initially not considered or visualized, because the primary objective is distance.
%\vspace{-1mm}

%it is important to offer a high quality of service in order to attract more customers and encourage them to use their network as much as possible, thus providing higher average revenue per user

\subsection*{Optimization Algorithm: Dragoon}
We developed a new optimization algorithm Dragoon %(Diversification Rectifies Advanced Greedy Overdetermined Optimization N-Dimensions)
for server locations.
%, based on the k-Means strategy \cite{MacQueen67,Lloyd82}.
The reference algorithms are described in Section \ref{algorithms}. To reduce the effect of the sensitivity to initially randomly selected server locations and to avoid multiple runs, we create a deterministic initialization and optimization algorithm. Usually, the first placed server serves a high amount of customers and has a large impact on the resulting structure. %This is possible in the planning phase, but it shows the significant influence of the first placement decision.
This can lead to impractical configurations if we consider limited capacities.
In these cases, an even distribution would be desirable to balance the load of servers. Dragoon consists of two main steps. After the initialization, the vertices are assigned to the nearest server location. In an iterative optimization these locations are improved.

In the preliminary stage of the initialization phase, an orientation mark is placed at the optimal location according to the one server placement problem.  % A complete search is necessary, but the calculation of the one server case can be done fast.
The first server is placed at the position of a network node which is farthest away from the orientation mark depending on network distance .
This mark is only for orientation to place the first server and is removed after the first location was identified. %to obtain uniform distributed nodes for initialisation. 
Subsequently, the remaining amount of server is planned using the adapted Two-Approx strategy. It calculates for every network node the distance to all servers with Dijkstra's algorithm. It chooses the location with the largest distance to its closest server as the additional server location. %Thereby, we obtain a very specific solution of the 2-Approx placement strategy.
%The orientation mark is removed after the first server location was identified.
%Thereby, the guarantee of the 2-approximable quality of the result \cite{Hochbaum85,Vazirani01} is still valid and it eliminates random influences.
%This ensures that the maximum value of  a distance from a network node to its closest Landmark is  not larger than twice the maximum considering the optimal placement location of all Landmarks.
After the initialization, the algorithm starts with the iterative refinement to recalculate and further optimize the server locations. These newly designed optimization steps are adaptable to different placement constraints.
%After initialization, the algorithms uses the classical k-Means approach. 

%Based on this knowledge, we developed a refinement rule to recalculate the centers with respect to the goal function. Using the divide and conquer principle, the 1-center problem for the node placement constrained can easily be solved nearly optimally for each group of customers and their center. 
The algorithm tests all possibilities of better server locations among all network nodes within one hop distance and direct connection from the current location in each iteration step. The new location is chosen according to the best improvement. The optimization steps are iteratively calculated for each server. The improved location is evaluated with respect to the entire scenario, because a single change influences the complete system. In every iteration step, the customers are reassigned to their nearest server. 
All actual server locations are used in every evaluation step except of the observed one. 
% If the new location improves the overall situation, the algorithm accepts it and replaces the current observed server location with the new one.
This is done with respect to the specified optimization criterion, the maximum distance calculated by Dijkstra's algorithm.

If this main fitness value is unchanged, the algorithm will use another additional criterion. To choose between two solutions and to identify an improvement, we use an average or mean criterion. In each iteration, every server is allowed to shift its location only once. This leads to a stepwise improvement and avoids a premature stagnation in a local optimum. The order of the server selection has no significant influence on the final result. This is due to the global view onto the problem. For our simulations, the servers are chosen with respect to the largest distance of assigned user first. This iterative optimization is repeated until all server locations do not change any more. Due to the described initialization, only a few iterations are necessary until the algorithm terminates.

%For the focused application area, the algorithm simply tests all locations of grouped vertices for a server. To identify an improved location, the algorithm evaluates the overall scenario. 
 
% 
The algorithm accepts only improved locations in every iteration step. Therefore, the 2-approximable condition of algorithm Two-Approx holds. Nevertheless, we show that the performance is much better, close to the global optimum. This optimization is calculated in polynomial runtime and it will always terminate. The algorithm can also be adapted to upgrade an existing scenario with partly fixed servers from the beginning or other constraints. A typical application area for this algorithm is the clustering of data.\\

\vspace*{-2mm}
\section{REFERENCE PLACEMENT ALGORITHMS}\label{algorithms}
%The algorithms are categorized into two classes, one for constraint free placement and one for node placement. Algorithms for free placement are also used for the infrastructure placement problems. We just include an additional optimization step where we translate the results by moving the generated server locations to the nearest infrastructure point. We do not consider algorithms for infrastructure-placement at the moment, since all free placement algorithms can easily be adapted to generate appropriate solutions. It is therefore sufficient to consider the most restrictive case (node placement) and the most flexible approach (free placement). 

The following algorithms are used as reference for our own development for the optimized placement without giving attention to customer profiles. These are specifically designed for the k-center problem, except for the evolutionary algorithms. All approaches can also be adapted to upgrade an existing scenario with partly fixed centers.

\begin{itemize} %[itemsep=2pt] 
\item Integer Linear programming
\item Two-Approx
\item Greedy
\item k-Means (MacQueen, k-Means++)
\item Evolutionary algorithm (SEREIN Framework)
\end{itemize}

\subsection{Integer Linear Programming}
The problem is defined with mathematical equations and can be solved with integer linear programming (ILP). % Here, a model of the problem 
%Kleinberg2013
A solution is calculated with support of ILP-Solver, for example GUROBI %\cite{Gurobi}. %
or CPLEX. % \cite{CPLEX}.
Both use an individual implementation of the Dual Simplex algorithm
%from \citet{Dantzig1963}
to obtain a solution in polynomial runtime even a totally unimodular matrix is given. The Simplex algorithm does not necessary calculate the global optimum due to internal model relaxation, transformation, and rounding.
The following ILP model %\cite{Pedroso2011}
is in accordance with a simple non capacitive problem, which has an integrality gap of 2 \cite{AaronArcher2003}. This guarantees a solution with fitness $d \leq 2 \cdot optimum$.

%Description of the model:
\begin{itemize} %[itemsep=2pt] 
\item The node $u_{j}$ is selected as location of a mirror server (1) or not (0), Equation \ref{LP1}.
\item The number of selected nodes correspond to the specified amount $k$ of mirror server, Equation \ref{LP2}.
\item A connection between two nodes $x_{ij}$ can exist (1) or not (0). This corresponds to the assignment of a customer to a defined mirror server, Equation \ref{LP3}.
\item Each node must have a connection to exactly one mirror server, Equation \ref{LP4}.
\item A customer can only be provided by a node, if this is selected as mirror server $c_{i}$, Equation \ref{LP5}.
\item Fitness function: The objective is the minimization of the largest distance. (dstc(ij) is the distance between two nodes) %\cite{Cygan2012}
, Equation \ref{LP7}.
\item Objective criterion, Equation \ref{LP9}.
\end{itemize}

\vspace*{-0.4cm}
\begin{figure}[H]
	\begin{minipage}{0.21\textwidth} 
\begin{equation}
u_{ j }\; \in \; {0, 1}\; \forall j
\label{LP1}
\end{equation}
	\end{minipage}
	\hfill
	\begin{minipage}{0.24\textwidth}
\begin{equation}
\sum_{ 0 }^{m}{u_{j}} \;=\; k
\label{LP2}
\end{equation}
	\end{minipage}
\end{figure}
\vspace*{-0.6cm}
\begin{figure}[H]
		\begin{minipage}{0.21\textwidth}
\begin{equation}
x_{ij}\; \in \; {0, 1}\; \forall j
\label{LP3}
\end{equation}
		\end{minipage}
\hfill
	\begin{minipage}{0.22\textwidth} 
\begin{equation}
\sum_{ 0 }^{n}{x_{ij}} \;=\; 1\; \forall j
\label{LP4}
\end{equation}
	\end{minipage}
\end{figure}
\vspace*{-0.7cm}
\begin{figure}[H]
	\begin{minipage}{0.21\textwidth}
\begin{equation}
x_{ij}\; \leq \; u_{j}\; \forall \; i, j
\label{LP5}
\end{equation}
	\end{minipage}
	\hfill
	\begin{minipage}{0.24\textwidth} 
\begin{equation}
dstc(ij)\; *\; x_{ij}\; \leq\; z \; \forall \; i, j
\label{LP7}
\end{equation}
	\end{minipage}
\end{figure}
\vspace*{-0.6cm}
\begin{figure}[H]
	\begin{minipage}{0.21\textwidth}
\begin{equation}
z \;\rightarrow\;min
\label{LP9}
\end{equation}
	\end{minipage}
 \hfill
\end{figure}
%\vspace*{0.1cm}

\subsection{Two-Approx}
%For this NP-hard k-center problem there exists a 2-approximable algorithm (Two-Approx).
Two-Approx guarantees a 2-approximable solution with cost d where  $d \leq 2  \cdot optimum$. Therefore the maximum value of a distance from a customer to his nearest mirror server is not larger than twice the maximum considering the optimal placement of the mirror servers \cite{Gonzalez85}. This guarantee is given without the knowledge of the optimum value. This algorithm is the best approximation we can get in polynomial run time and it is well studied \cite{Hochbaum85}. %,Vazirani01}. %We implemented the 2-approximable algorithms as our main reference, because of the known performance.
With these results, we are able to estimate the global optimum.

%The Two-Approx works as follows.
At the beginning, the algorithm choose a random node, which becomes the location of the first server. After that, it calculates for every node the distance to all other nodes. It chooses the node with the largest distance to their closest server as new location of the next server. This routine is repeated until the specified quantity of mirror servers are reached. %2-Approx places the mirror server as far away from each other as possible. Figure \ref{2-Approx-2Steps} shows the first 2 steps of the algorithm. On the left side the first server is randomly placed. In the next step, the algorithms places the second server as far away as possible.\\
% In comparison to the following Greedy strategy, the 2-approximable algorithms do not give attention on selected location for storage servers during runtime.
% So, the better choice is to place them as much as possible far away from each other.

%\begin{figure}[htb]
%{
%\centering
%\includegraphics[width=0.5\textwidth]{fig/2-Approx-2Steps.png}
%\caption{The first two steps of the 2-Approx.}
%\label{2-Approx-2Steps}
%}
%\end{figure}

\subsection{Greedy}
The Greedy strategy \cite{Jamin01} iteratively places the mirror servers on the topology at the customer locations. During each iteration, it tries all different placement possibilities for the next server and ultimately selects the node that provides the biggest benefit with respect to the optimization criterion. The Greedy strategy repeats this process until all mirror servers are placed. 

%For the implementation we use a distance matrix to determine the optimal placement and use an additional logic to avoid placing two or more servers at the same position. For the simple implementation the matrix is not updated for each optimization step. The extended approach updates the matrix after every placement. To improve the quality further we include backtracking, to test whether already placed servers can be placed better or may be removed completely.

%Table \ref{tab:Greedy} shows the distance matrix. For symmetric distances only the upper diagonal matrix needs to be considered. In this case, the location of Node N3 will be selected for the next mirror server.
%\begin{table} [htbp]
%\begin{center} 
%\caption{Example distance matrix of Greedy for 5 nodes. }
%\label{tab:Greedy}
%\begin{tabular}{|c|c|c|c|c|c|c|}
%\hline
%	& N1  & N2 & N3 & N4 & N5 &  \textbf{max} \\
%\hline
%\hhline{--}
%N1 &  -  & 5   & 10 & 13 & 8   & 13\\
%\hline
%N2 &  5 & -   & 4 & 17 & 9   & 17\\
%\hline
%\textbf{N3} & 10 & 4   & - & 12& 6   & \textbf{12}\\
%\hline
%N4 & 13 & 17  & 12 & - & 15   & 17\\
%\hline
%N5 &  8  & 9   &  6 & 15 & -   & 15\\
%\hline
%\end{tabular}
%\end{center}
%\end{table}

\subsection{k-Means}
% are adapted for our placement condition.
The main idea of the following clustering algorithms is to define \textit{k} center points, one for each cluster. In our case a cluster is a group of nodes. %Since placements of servers obviously influence the optimal locations of other servers, this approach places all center nodes at the same time. %This makes backtracking unnecessary.
The algorithms place all center nodes at the same time.
Various approaches %for clustering
exist for a fixed number of clusters, they differ mainly with regard to the initial placement of center points. The MacQueen \cite{MacQueen67} algorithm is one of the simplest k-Means algorithms. It relies on randomly selected locations, which are used as the initial locations of the mirror servers. % In contrast the algorithm from Lloyd \cite{Lloyd82}, also known as Voronoi relaxation, starts with completely randomly placed servers. 
Another typical initialization is used in the k-Means++ algorithm \cite{Arthur07}. Here, the location of the first server is chosen randomly at a node. The other servers are also placed randomly at a location of a node, however the probability is skewed to favor certain locations. The selection probability for locations is increased proportionally with their squared distances to already selected locations. 

After the initialization, the customers are assigned to their respective server. For each group of customers related to a shared server an updated server location is calculated. 
The positions of the server nodes are mapped to the nearest node either in every step or at the end of the optimization.
%The new center node is the geometrical center of all customers in a group.
This process is repeated until server locations do not change any more.
%\begin{equation}
%m_{ i }^{ t+1 }= \frac{ 1 }{ \left \lvert S_{ i }^{ t } \right \rvert } \cdot \sum{ x_{ j } \in S_{ i }^{ t } }{  }{ x_{ j } }
%\label{kmeans}
%\end{equation}
%After we have these new centroids, a new binding has to be done between the same data set points and the nearest new centroid.
After every iteration step, the customers are reassigned to the nearest mirror server. % It is proven that the procedure will always terminate. % \cite{Selim84}.
The algorithm is sensitive to the initially randomly selected mirror server locations and does not necessarily find the optimal solution. To reduce this effect, the algorithms are run multiple times. %To adapt this strategy from the free placement to the node placement, the positions of the center points are mapped to the nearest customer location either in every step or at the end of the optimization.
%\vspace{1mm}

\subsection{Evolutionary Algorithms}
%Finally, we use an evolutionary algorithm to optimize the server placement. 
The SEREIN framework \cite{serein} is used to implement the evolutionary algorithms. We employ the standard implementation of a genetic algorithm (GA) provided by SEREIN
and use a population with 25 individuals evolving over 80
generations. In addition, a particle swarm optimization (PSO)
and simulated annealing (SA) approach are implemented as
well. The parameters for the algorithms were determined
experimentally using meta-optimization.

%Finally we use an evolutionary algorithm to optimize the server placement. The SEREIN framework \cite{serein} is used to implement the algorithm. We employ the standard implementation of a genetic algorithm provided by SEREIN and use the following settings:
%\begin{itemize}
% \item Genetic representation - a vector of real numbers $r$ with $0 \leq r < 1.0$, that was translated into coordinates of the servers
% \item Variation operators - Creep Mutation with standard deviation 0.01 and no recombination operator
% \item A population with 30 individuals evolving over 80 generations
% \item Tournament selection with tournament size 2
%\end{itemize}

\vspace{1mm}
\section{SIMULATION AND ASSESSMENT}\label{simulation}
For the simulation of the model, the experiments, and the entire system we use a prototypic implementation in Java. For the evaluation, we run multiple simulations and compare different optimization algorithms. We used classical geo-coordinates in the 2-dimensional space and the described hop distance combined with euclidean distance as metric. % A scenario can be stored to and retrieved from an XML file. The customers, the servers, and the assignments of customers to their nearest servers of a scenario are visualized using the data format GeoJson \cite{geojson} in TileMill \cite{tilemill}.
For the evaluation, the fitness function calculates the distance parameters: maximum, quantiles, median and average. To generate significant results, we performed repeated optimizations using five different network topologies from \textit{the Internet topology zoo} \cite{Knight2011}. All used real-world scenarios have a similar amount of backbone network nodes, approximately 100.
% Table \ref{test-setups} show the parameter of the test setups.
The presented results are the average values of all scenarios for each algorithm.

% Simulationen
% 1. Ziegen wie gut man ist gegen LP (Scenarien mit LP und Dragoon) + 2-Approx + Theoretical Optimum
% 2. Notwendige Anzahl an Servern + Große Tabelle
% 3. Einfluss von Caching

\begin{figure}[htb]
 {
 \setlength{\abovecaptionskip}{0ex}%
 \setlength{\belowcaptionskip}{0ex}%
 \centering
 \includegraphics[width=0.44\textwidth]{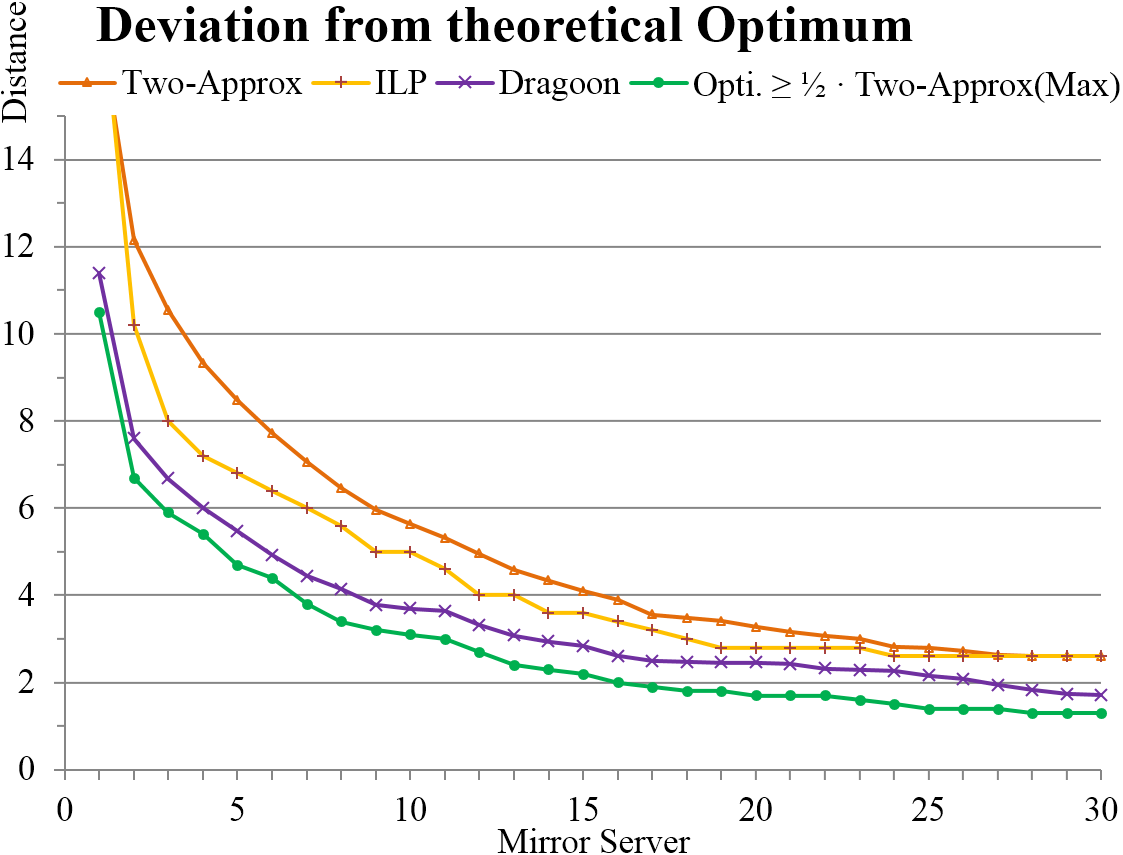}
  \caption{Comparison of Dragoon, Two-Approx, and ILP with the theoretical Optimum based on maximum distances.}
 \label{TheoreticalOptimum}
 \vspace*{-0.2cm}
 }
\end{figure}

 % 1---------------------------------- Gegen LP
Initially, the performance of our developed Dragoon algorithm is compared with the approaches of Two-Approx and ILP. Since the solutions of these reference algorithms have a proven performance, the results serve as basic benchmark. The solutions of Two-Approx vary because of the random initialization. Nevertheless, the 2-approximable condition is valid every run. Based on the results $s$ of the Two-Approx and ILP we defined a theoretical limit for the optimum. %, because the real optimum is unknown due to the NP hardness of the problem:
 \vspace*{-0.1cm}
\begin{equation}
 s \leq 2 \cdot Optimum_{ real} \implies \frac{ 1 }{ 2 } \cdot s = Optimum_{ theoretical}
\end{equation}
Since the 2-approximable condition is always valid, the highest values of the multiple test runs can be used for the estimation of the theoretical optimum instead of the average values.
Figure \ref{TheoreticalOptimum} shows that Dragoon returns much better results than the other approaches. The Dragoon algorithm reaches a performance close to the estimated theoretical optimum. For larger server amounts, the quality of solutions and the improvement with additional servers stagnates for all algorithms. % with their performance nearly,
It indicates that we are already close to the optimum.
\vspace*{-0.1cm}
 % 1 ENDE ---------------------------------- Gegen LP

%Figure \ref{DeviationToTheorecticalOptimum} shows the results of algorithm comparison. For small server numbers, our improved Conny algorithm is close to the theoretical optimum. Also the evolutionary heuristic of the framework SEREIN achieves very good results. For larger server counts both algorithms stagnate with their performance nearly at the same level, indicating that we are already very close to the actual optimum. Both approaches performed significantly better than 2-Approx and were much faster than a brute force approach.

%  \begin{figure}[htb]
%   {
%   \centering
%   \includegraphics[width=0.5\textwidth]{fig/DeviationToTheorecticalOptimum.png}
%   \caption{Deviation between 2-Approx and our best algorithms for maximum distance.}
%   \label{DeviationToTheorecticalOptimum}
%   }
%   \end{figure}

% 2. Notwendige Anzahl an Servern + Große Tabelle
Based on the maximum distance of a customer to its nearest mirror server, Table \ref{tab:PerformanceComparisonMax} and Figure \ref{fig:PerformanceComparisonMax} present the general Performance of the different algorithms. It shows that it is sufficient to set up a server at about 15 \% of the nodes. After we placed 17 mirror server for about 110 network nodes, the average improvement of maximum distance for an additional server is less than 1 \% with the Dragoon algorithm in all scenarios.
While additional servers have a positive effect, increasing the overall capacity and load balance, the added benefit decreases significantly for large server counts. For high ratios of servers in relation to customer count we observe a saturation effect.
The performance of the SEREIN framework with evolutionary algorithms is remarkably. SEREIN is not customized to this problem but reaches a comparably good solution as the specifically designed algorithms. The parameters of the heuristic were adapted after as little as six small test runs. The calculation time of all algorithms depends strongly on their parameters and predefined time limits. The time complexity of the algorithms is considerably different, but all optimization runs finished after a couple of minutes or a few hours. For such a fundamental decision as the placement of mirror servers, the computation time is acceptable. Decisions have strong long term effects and directly influence future service quality and costs.
% We also analyzed this aspect for different customer quantities. The direct number of servers is nearly the same. So we know that this is not directly dependend from the number of customers. It depends also on the size of the area (bounding box), where the customers are placed and the degree of satiation level.
%\vspace*{-0.1cm}
\begin{table} [htbp]
\setlength{\belowcaptionskip}{0ex}%
\begin{center} 
\caption{Performance overview with objective maximum distance in relation to amount of mirror servers.}
\label{tab:PerformanceComparisonMax}
\setlength{\tabcolsep}{4pt}
\begin{tabular}{|c!{\vrule width 0.8pt}c|c|c|c|c|c|c|c|c|c|c|c|c|c|}
\hline
\rotatebox{90}{Server} & \rotatebox{90}{\textbf{Dragoon}} & \rotatebox{90}{Two-Approx} & \rotatebox{90}{ILP} & \rotatebox{90}{Greedy} & \rotatebox{90}{Greedy BT}  & \rotatebox{90}{GRASP} & \rotatebox{90}{k-means} & \rotatebox{90}{Serien GA} & \rotatebox{90}{Monte Carlo} & \rotatebox{90}{$Opti._{ theoretical}$} \\ \hline
\hhline{----------}
%1      & Dragoon    & 2-Approx           & ILP         & Greedy & Greedy Bt & GRASP &kmeans          & GA           & Monte Carlo         \\ \hline
1		& \textbf{11,4}	& 	17,2			&	18,4		&	11,4 &	11,4		& 11,4 & 11,4		& 12,6			& 14,9		& 10,5 \\ \hline
2      & \textbf{7,6}   &    12,2           &   10,2        &   11,2  & 8          &  8  		& 7,7	     & 8,8            & 10,5      &  6,7 \\ \hline
3      & \textbf{6,7}   &   10,5            &   8        	 &  8,4      & 7,4        & 7  		 & 7,4	      &  8,2            &  9          &   5,9 \\ \hline
5      & \textbf{5,5}    &     8,5          &   6,8         &   7,6       & 6,2     &  6,4 		& 6,2	     & 6,8            & 7,8        &   4,7 \\ \hline
10     & \textbf{3,7}   &     5,6          & 5              &   5,8       & 4,2       & 4,8 	&	5,1	  & 5,4            & 6,3         &   3,1 \\ \hline
15     &  \textbf{2,8} 	&  4,1            & 3,6        	 &   5           & 3,8  & 4,4  			&  4,3       &  4,8            &  5,1       & 2,2 \\ \hline
20     & \textbf{2,5}   &    3,3          & 2,8          &    4,2       & 3,2     & 3,8  	  & 3,9	     & 4,2             & 4,5        &  1,7 \\ \hline
30     & \textbf{1,7}   &    2,6         & 2,6            &    4         & 2,8   & 3,2  		& 2,9	       & 3               & 3,8        &  1,3 \\ \hline
\end{tabular}
\end{center}
\vspace*{-0.1cm}
\end{table}

%\vspace*{-0.1cm}
\begin{figure}[htb]{
\setlength{\abovecaptionskip}{0ex}%
\setlength{\belowcaptionskip}{0ex}%
\centering
\includegraphics[width=0.46\textwidth]{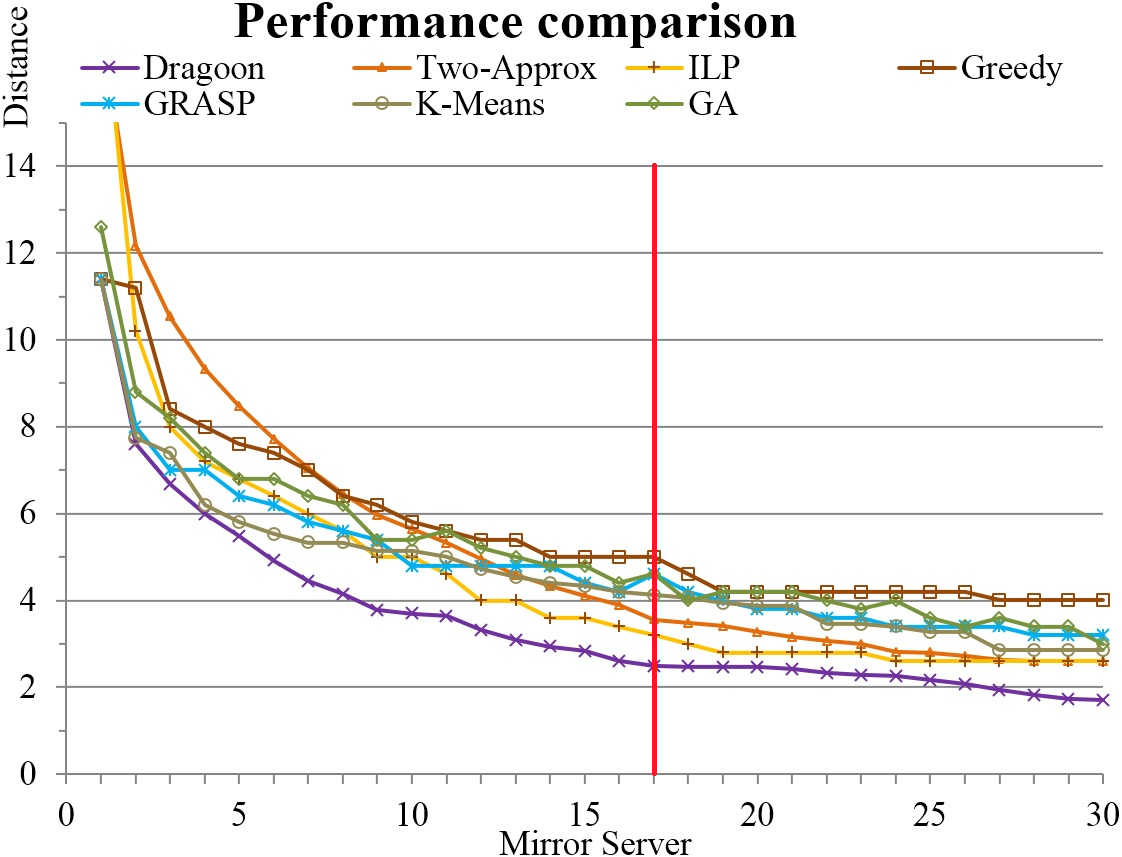}
\caption{Performance comparison of different algorithms. The red vertical line marks the quantity of mirror server, after which the average performance improvement is less than 1 \%.}
\label{fig:PerformanceComparisonMax}
\vspace*{-0.5cm}
}
\end{figure}
% 2. ENDE : Notwendige Anzahl an Servern + Große Tabelle

A mirror server should have all requested data in its cache to reply immediately. The server caches mainly reactive and loads missing data from other sources at request time.
Therefor, the customer requested content on a server shall be positive correlated so that the mirror server obtain requests for current available data. This increase the cache hit ratio on a server, lowers the network utilization and reduces access latency. Furthermore, it avoids redirecting to other servers and lowers the load on the original source.
In the following Experiment we compare different optimization objectives. On the one hand, the optimization has been in terms of minimizing the distance between the customers and mirror servers, regardless of the customer profiles. On the other hand, it has been optimized with regard to a positive correlation of customer profiles at the mirror server. To optimize the profile correlation, we used the Spearman's rank correlation coefficient. % \cite{Spearm1904}.
Customers are reassigned after an initial placement optimization to the best fitting mirror server. Following, updated locations of the servers are calculated based on the assignment. Table \ref{tab:BenachmarkCache} shows the comparison of the two conflicting objectives.
\vspace*{-0.1cm}
\begin{table} [htbp]
\begin{center} 
\caption{Benchmarking with respect to different optimization objectives. }
\label{tab:BenachmarkCache}
\setlength{\tabcolsep}{4pt}
\begin{tabular}{c|c|c}
%\hline
 & Objective: Distance & Objective: Profile \\ \hline
Average request distance & 1,6 Hops & 6,1 Hops  \\ \hline
Cache Miss Ratio & 36 \% & 5 \%  \\
\end{tabular}
\end{center}
\vspace*{-0.2cm}
\end{table}
    
%\vspace{1mm}
\section{CONCLUSION AND OUTLOOK}
%Our study showed a significant ..
In this paper, we propose an adaptive model for the placement problem of mirror server.
We have developed a new placement strategy, which outperforms the others and finds solutions close to the optimum in short time. To enhance the QoE of customers and the access performance, we analyzed the quantity of recommended mirror servers for a predefined area. After a specific number of servers is placed for a CDN, it is more important to improve the cache hit ratio than to further reduce the distance and access latency with additional mirror servers. Our study showed a significant cache hit improvement by optimizing positive correlated customers profiles instead of just reduced network distances. The trade-off between these contradictory objectives has to be taken into account by the management of the CDN provider. For an effective and balanced interim solution much more effort is necessary.
%With this work, we provide supporting results.
If we add more mirror server than 15\% of the number of network nodes, we reached a saturation effect.
Our analysis shows, even for the best placement strategy less than 1 \% performance gain can be expected per additional server. This is important for management decisions between adding new mirror servers or extending existing ones. With improved performance of the placement algorithms, we can further decrease the number of server locations. In the future, we intend to further enhance the performance of the placement algorithms. Another open question is the optimal distance to the second closest server to provide a high fault tolerance. We are working on an optimization, which reaches good solutions for both objectives at one time, reduced distance and high cache hit ratio.\\

%\vspace*{-0.4cm}
%\section*{Acknowledgment}
%This work was partly funded by FLAMINGO, a Network of Excellence project (ICT-318488) supported by the European Commission under its Seventh Framework program.
%\vspace*{-0.2cm}

%\balance
%\bibliographystyle{IEEEtran}
\bibliographystyle{unsrtnat}
\bibliography{IEEEabrv,literature}

\end{document}